\documentclass{PoS}
\usepackage{caption}
\usepackage{subcaption}
\usepackage{url}
\usepackage{lineno}
\newcommand{\doublefig}[9]{
	\begin{figure*}[!]
		\begin{subfigure}[t]{0.49\textwidth}
			\fbox{\includegraphics[width=\textwidth,height=#1\textwidth]{#2}}
			\caption{{\small #3.}}
			\label{#4}
		\end{subfigure}
		\begin{subfigure}[t]{0.49\textwidth}
			\fbox{\includegraphics[width=\textwidth,height=#1\textwidth]{#5}}
			\caption{{\small #6.}}
			\label{#7}
		\end{subfigure}
		\caption{{\small #8.}}
		\label{#9}
	\end{figure*}
}
\newcommand{\onefig}[5]{
	\begin{figure}[!]
		\centering
		\fbox{\includegraphics[width=#1\textwidth,height=#2\textwidth]{#3}}
		\caption{{\small #4.}}
		\label{#5}
	\end{figure}
}
\newcommand{\esqtab}[6]{
	\begin{table}
		\begin{center}
			\scalebox{#3}[#4]{
				\begin{tabular}{#5}
					#6
				\end{tabular}
			}	
			\caption{{\small #1.}}
			\label{#2}
		\end{center}
	\end{table}
}
\title{Stability and behavior of the outer array of small water Cherenkov detectors, outriggers, in the HAWC observatory}
\ShortTitle{Stability and behavior of the array of outriggers in the HAWC observatory}
\author{\speaker{T. Capistr\'an}$^a$, I. Torres$^a$, E. Moreno$^b$ and for the HAWC Collaboration$^c$ \\
        \llap{$^a$}Instituto Nacional de Astrof\'isica, \'Optica y Electr\'onica, Luis Enrique Erro 1, Tonantzintla, Puebla 72840, M\'exico \\
	\llap{$^b$}Facultad de Ciencias F\'isico Matem\'aticas, Benem\'erita Universidad Aut\'onoma de Puebla, Ciudad Universitaria, Colonia San Manuel Puebla, M\'exico \\
        \llap{$^c$}For a complete author list, see \href{http://www.hawc-observatory.org/collaboration/icrc2017.php}{http://www.hawc-observatory.org/collaboration/icrc2017.php}.\\
        Email: \email{tcapistran@inaoep.mx}, \email{ibrahim@inaoep.mx}, \email{emoreno@fcfm.buap.mx}}

\abstract{The High-Altitude Water Cherenkov (HAWC) Observatory is used for detecting TeV gamma rays. HAWC is operating at 4,100 meters above level sea on the slope of the Sierra Negra Volcano in the State of Puebla, Mexico, and consists of an array of 300 water Cherenkov detectors (WCDs) covering an area of 22,000 $m^2$. Each WCD is equipped with four photomultiplier tubes (PMTs) to detect Cherenkov emission in the water from secondary particles of extensive air-shower (EAS) that are produced in the interactions of primary particles (gamma rays or charged cosmic rays) in the atmosphere. HAWC is able to reconstruct the EAS in the 0.5 to 100 TeV energy range. In order to improve the core determination for events with high energy ($>10$ TeV) when the events arrive outside of the HAWC array, the Outrigger upgrade project is adding 350 small WCDs around the main array. These outrigger tanks each have one PMT in a 1.5 meter diameter cylindrical polyethylene tank, covering a total area four times larger than that of the HAWC array. In this work we present leak light testing to identify the stability of the detector and an analysis of deposited charges to understand the detector performance.}

\FullConference{35th International Cosmic Ray Conference --- ICRC2017\\
		10--20 July, 2017\\
		Bexco, Busan, Korea}

\begin{document}
	\section{Introduction}
		\paragraph*{}The High-Altitude Water Cherenkov (HAWC) observatory is located at an elevation of 4,100 meters above sea level on the flank of the Sierra Negra volcano in the state of Puebla Mexico. HAWC consists of an array of 300 water Cherenkov detectors (WCDs). The HAWC WCDs are metal cylindrical structure of 7.3m in diameter and 4.5m high, lined with a dark plastic bladder, and holding nearly 200,000 liters of high-purity water and, on the bottom, four photomultiplier tubes (PMTs). The array covers an area of 22,000 $m^2$ that is able to detect primary particles in the energy range from 500 GeV to 100 TeV \cite {2017arXiv170407411A, 2017ApJ...841..100A, 2017arXiv170101778A}.
		\paragraph*{}When a primary particle impinges on the Earth's atmosphere, it starts interacting with the atoms of the air and produces an cascade of secondary particles. These secondary particles can penetrate the WCDs, and emit Cherenkov light that is detected by PMTs. The footprint of an event on the ground mainly depends on the energy of the primary particle. When it has a high energy, the most PMTs in the array are activated. However, if the event's core falls outside the array, it may be poorly reconstructed \cite{2015arXiv150904269S}. Therefore an array of small water Cherenkov detector, called outriggers, will be installed around the HAWC's array (See Figure \ref{Fig:ORarray}). An outrigger is composed of a cylindrical polyethylene tank with approximate 1.5 meters in diameter and 1.4 meters high, filled with 2,000 liters of water and equipped with a single PMT at the bottom. The outrigger array will improve the core reconstruction and increase 3-4 times the sensitivity of HAWC to gamma-rays above 10 TeV \cite{2015arXiv150904269S}. In this work we present a test for identifying the stability of first set of outrigger detectors deployed at the site, and in order to understand the behavior of the outrigger, an analysis of charge distribution was done.
			\onefig{0.6}{0.5}{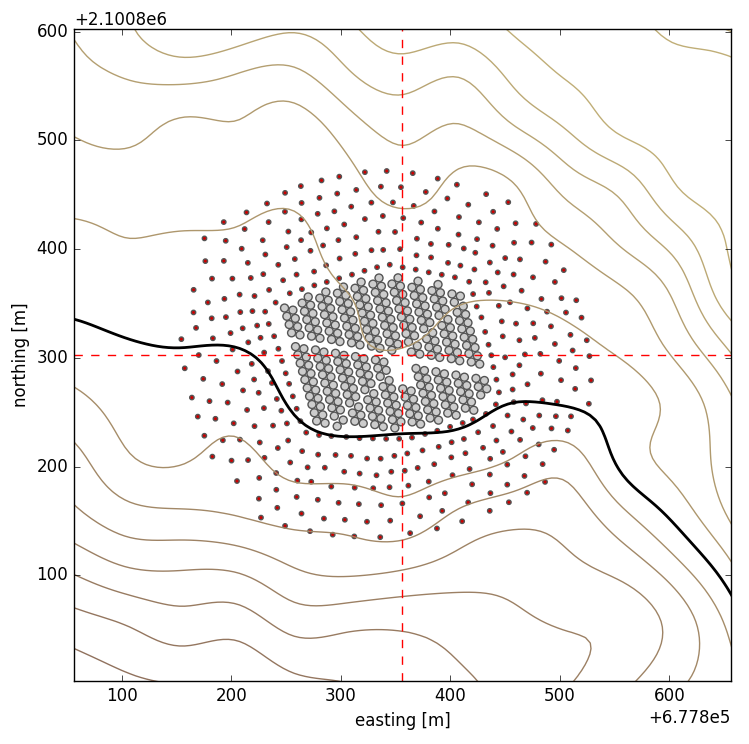}{Layout of the HAWC central and outrigger arrays. The unit of the axis are in UTM that belongs to zone 14}{Fig:ORarray}
	\section{Experimental setup}
		\paragraph*{} In Figure \ref{Fig:setup} shows one diagram that illustrates the flow of the information from PMT signal to data acquisition, This diagram is a typical setup for collecting information about photon counting and signal integration \cite{2017NIMPA.861...28G,ALARCON199939}. Each outrigger has one Hamamatsu R5912 8" PMT at the bottom and is powered by a CAEN V6533 high voltage VME module at an optimal tension (see Table \ref{Fig:OR}). A decoupler was used that is a capacitive pickoff circuit to take the PMT AC signal off of a DC HV. Figure \ref{Fig:pulse} shows a typical PMT signal. Two types of measurement were taken:  the rate (the number of PMT signal per second) and PMT signals.
			\onefig{0.8}{0.35}{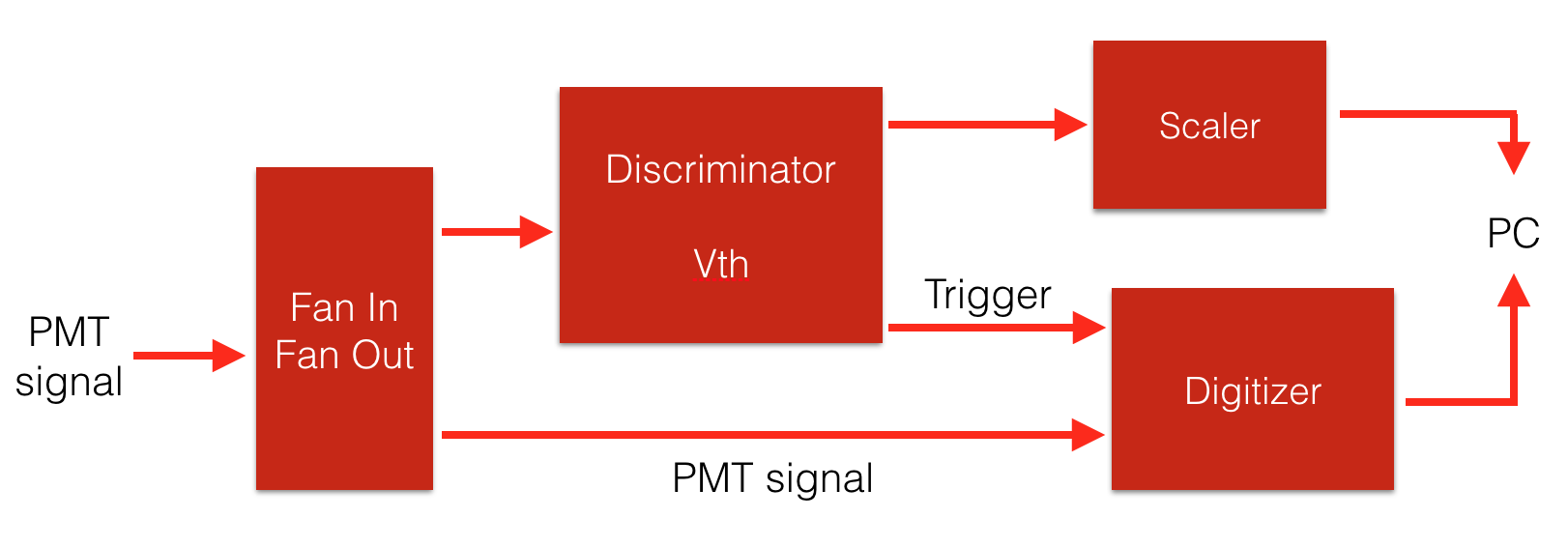}{Signalling block diagram for the outrigger PMTs}{Fig:setup}		
		\paragraph*{}To take the rate was used two modules: discriminator (CAEN V812) and scaler (CAEN V820). The discriminator produces an output pulse when the signal is greater than the threshold voltage (Vth), in this case, Vth is -3 mV. The scaler module counts the numbers of discriminator output pulse during a windows time of one second and is saved a text file.
		\paragraph*{}The PMT signal is digitized with a modulation frequency of 1 GHz during a time windows of 800 ns using a CAEN V1751 digitizer module. Only the PMT signals that have a voltage greater than the Vth were stored. The Vth is set -3 mV in the discriminator module.
			\onefig{0.7}{0.4}{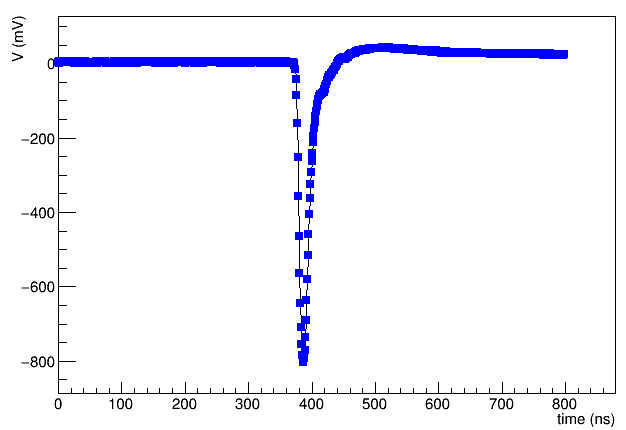}{Digitized signal from a PMT}{Fig:pulse}
	\section{Meansurements and results}
		\paragraph*{}The light leak is one problem with these detectors due to they are extremely sensitive. The rate may have measurements in order of MHz or a lot of fluctuations during data acquisition if the outrigger has a light leak. In this work, we took the rate measurement of eight outriggers at the same time during 20 hours from 5 pm to 1 pm (Mexico City time). Figure \ref{Fig:za11} shows the rate of ZA1-1 outrigger during 20 hours. It is stable because the rate is around 10 kHz regardless of the weather conditions, therefore, the outrigger is correctly sealed. In order to compute the mean rate, a histogram was made with all rate measurement  and a Gaussian function was fitted (Figure \ref{Fig:fit}). However, the ZA1-4 outrigger has a hole that permits the pass of the light into the tank because the graph of the rate (Figure \ref{Fig:za14}) has several peaks around 16:00 UTC (11:00 am Mexico City time) when the weather was sunny before it was cloudy. The hole was detected using a lamp that was lighted around the tank when it is close to the hole, the rate has an increase. 
The outriggers, that were found this issue, were fixed.  Ideally, we expect all outriggers have the same outcome without important of the voltage that was fed. In the Table  \ref{Tab:rate}, the rate of most outriggers has a value around 10 kHz but there is one that its rate is around 13 kHz (maybe the PMT gain is higher than other).
			\doublefig{0.8}{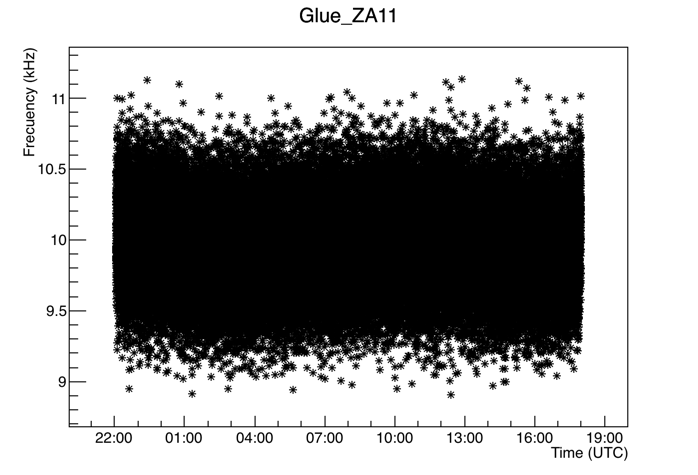}{Rate of ZA1-1 outrigger}{Fig:za11}{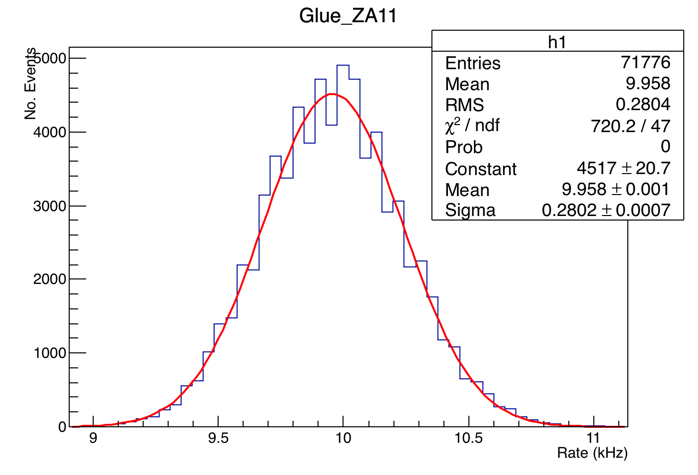}{The Gaussian function is fitted at the histogram of the rate}{Fig:fit}{In (a) shows a stable rate because it does not have a lot of fluctuations during the data acquisition. With the rate that was taken during around 20 hours, a histogram was made and a Gaussian function was fitted in order to compute the mean rate. This histogram is shown in (b)}{Fig:OR}
			\onefig{0.6}{0.4}{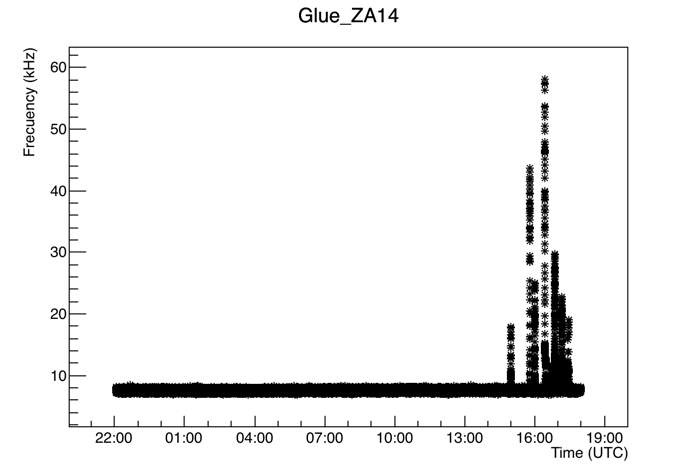}{A graph of the rate of ZA1-4 outrigger. Around 16:00 hours, there are several peaks because in the tank has a leak}{Fig:za14}
			\esqtab{The operation voltage that was fed of each outrigger, and mean of the rate of the Gaussian fit}{Tab:rate}{1}{1}{|c|c|c|}{
				\hline
				Outrigger's name & Voltage & rate (kHz) \\
				\hline
				ZA1-1 & 1455 & 9.96 \\
				\hline
				ZA1-2 & 1598 & 8.91\\
				\hline
				ZA1-4 & 1497 & 7.64\\
				\hline
				ZA1-5 & 1547 & 12.79\\
				\hline
				ZA1-6 & 1798 & 9.69\\
				\hline
				ZA1-7 & 1659 & 10.69\\
				\hline
				ZA1-8 & 1765 & 9.62\\
				\hline
				ZA2-3 & 1617 & 8.10\\
				\hline
			}
		\paragraph*{} In order to check the variation of the charge at different times, 50,000 pulses (PMT signals) were saved each 6 hours in one outrigger, then the bunch of pulses was taken at afternoon, evening and morning (the time in UTC is shown in the Table \ref{Tab:fitgauss}). Each pulse was integrated for obtaining the histogram of charge (see figure \ref{Fig:histo}), the PMT pulses that has a voltage equal or greater than 1 V is avoided, therefore in the histograms has fewer events for obtaining the histogram of charge. The figure \ref{Fig:histosolo} show the histogram of charge of the first 50,000 pulses. According to \cite{Etchegoyen:2005xg}, in this type of histogram has two peaks: the first one corresponds to low charge values is deposited by short-tracked muons and other particles (gamma after conversion, electrons, and electronic noise), and the second one corresponds to background muons. The other bunch of pulses help to check the check the stability of the gain of the outrigger due to their histogram of charge (figure \ref{Fig:histogrupo}) looks like similar and does not change when it is at different weather condition, that is, there are approximate the same number of pulse with the same charge in each histogram. In the range from 80 to 150 pC of these histograms are located the second peak, a Gaussian function was fitted, and we export the mean and sigma, and included in the Table \ref{Tab:fitgauss}. the values of these fits are around 107 pC, that it means the PMT gain is stable at different weather condition.		
			\doublefig{0.8}{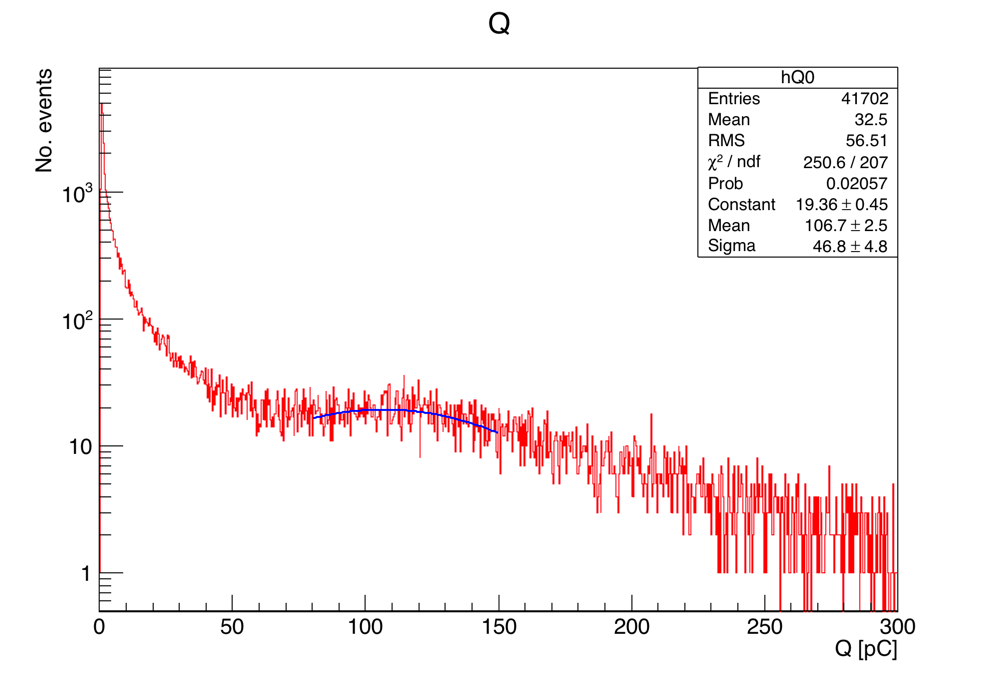}{Histogram of charge}{Fig:histosolo}{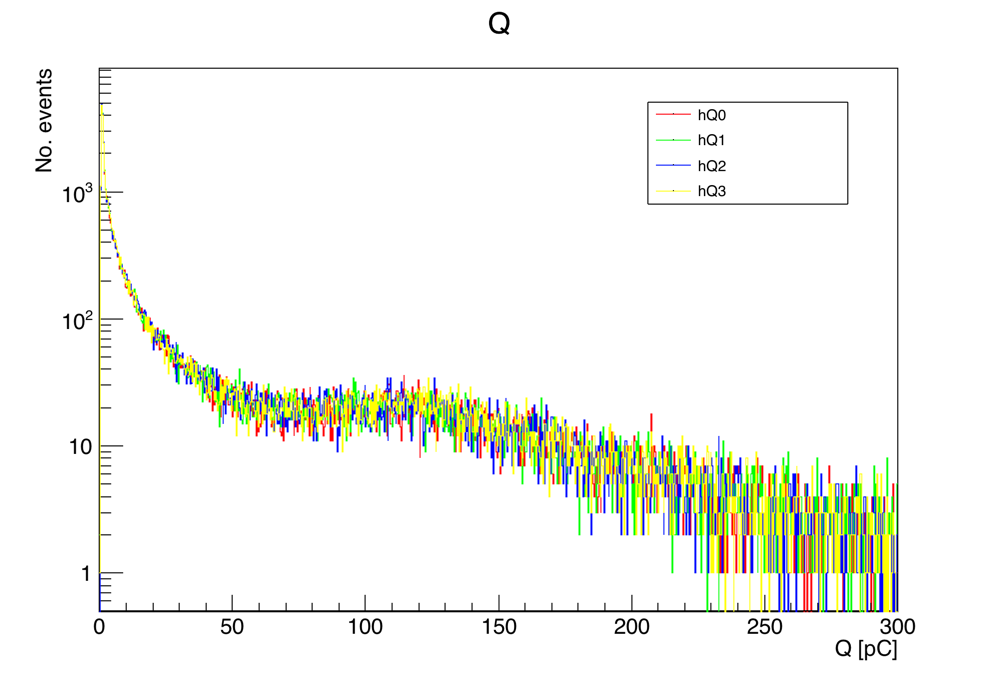}{Histogram of charge at different time}{Fig:histogrupo}{In one outrigger was taken 50,000 pulses each 6 hours, these pulses were integrated in order to produce the histogram of charge. In (a) shows the integration of the first 50,000 pulses that was taken (Q0). There is a small peak in the range from 80 to 150, then a Gaussian function was fitted. In (b) shows the histogram in (a) and other three histograms. They do not present any variation each other due to the morphology is the same. In these histograms, a Gaussian function was fitted, the mean and sigma was exported and included in the Table \ref{Tab:fitgauss}}{Fig:histo}
			\esqtab{In the histogram of charge of the Figure \ref{Fig:histo} was fit a Gaussian function from 80 to 150 nC and export the mean and sigma. Also, it shows the time that started to collect data}{Tab:fitgauss}{1}{1}{|c|c|c|c|}{
				\hline
				Name & time (UTC) & mean & sigma \\
				\hline
				Q1 &  22:00 & 106.7 & 46.8\\
				\hline
				Q1 & 04:00 & 104.8 & 49.8\\
				\hline
				Q2 & 10:00 & 107.1 & 42.4\\
				\hline
				Q3 & 16:00 & 110.2 & 40.9\\
				\hline
			}
	\section{Conclusion}
		\paragraph*{}In this work, the light leak test applied to eight outriggers in order to identified the light leak. In the outrigger that has any issue, the rate increase is around 500 \%, that is, the normal rate is approximate 10 KHz, if there is a hole, it is 60 kHz. With small modifications to the outrigger detectors these can most likely be avoided during the full deployment of the outrigger array in the coming months.
		\paragraph*{} Four bunches of 50,000 pulses were taken at different time. The pulses were integrated to obtain the histogram of charge. The morphology of these histograms does not change. In each histogram, there is a small peak in the range from 80 to 150 pC. From the fitted mean of the Gaussian, we learn that in the time period that the 4 data sets were taken the gain of the detector was roughly constant. 
	\section*{Acknowledgments}
		\footnotesize{We acknowledge the support from: the US National Science Foundation (NSF); the
US Department of Energy Office of High-Energy Physics; the Laboratory Directed
Research and Development (LDRD) program of Los Alamos National Laboratory;
Consejo Nacional de Ciencia y Tecnolog\'{\i}a (CONACyT), M{\'e}xico (grants
271051, 232656, 260378, 179588, 239762, 254964, 271737, 258865, 243290,
132197), Laboratorio Nacional HAWC de rayos gamma; L'OREAL Fellowship for
Women in Science 2014; Red HAWC, M{\'e}xico; DGAPA-UNAM (grants IG100317,
IN111315, IN111716-3, IA102715, 109916, IA102917); VIEP-BUAP; PIFI 2012, 2013,
PROFOCIE 2014, 2015;the University of Wisconsin Alumni Research Foundation;
the Institute of Geophysics, Planetary Physics, and Signatures at Los Alamos
National Laboratory; Polish Science Centre grant DEC-2014/13/B/ST9/945;
Coordinaci{\'o}n de la Investigaci{\'o}n Cient\'{\i}fica de la Universidad
Michoacana. Thanks to Luciano D\'{\i}az and Eduardo Murrieta for technical support.
		}
	\bibliographystyle{JHEP}
	\bibliography{biblio}

\providecommand{\href}[2]{#2}\begingroup\raggedright\begin{thebibliography}{1}

\bibitem{2017arXiv170407411A}
A.~U. Abeysekara et~al., {\it {The HAWC real-time flare monitor for rapid
  detection of transient events}},  {\em ArXiv e-prints} (Apr., 2017)
  [\href{http://arxiv.org/abs/1704.7411}{{\tt arXiv:1704.7411}}].

\bibitem{2017ApJ...841..100A}
A.~U. Abeysekara et~al., {\it {Daily Monitoring of TeV Gamma-Ray Emission from
  Mrk 421, Mrk 501, and the Crab Nebula with HAWC}},  {\em The Astrophysical
  Journal} {\bf 841} (June, 2017) 100,
  [\href{http://arxiv.org/abs/1703.6968}{{\tt arXiv:1703.6968}}].

\bibitem{2017arXiv170101778A}
A.~U. Abeysekara et~al., {\it {Observation of the Crab Nebula with the HAWC
  Gamma-Ray Observatory}},  {\em ArXiv e-prints} (Jan., 2017)
  [\href{http://arxiv.org/abs/1701.1778}{{\tt arXiv:1701.1778}}].

\bibitem{2015arXiv150904269S}
A.~{Sandoval}, {\it {HAWC Upgrade with a Sparse Outrigger Array}},  {\em ArXiv
  e-prints} (Sept., 2015) [\href{http://arxiv.org/abs/1509.4269}{{\tt
  arXiv:1509.4269}}].

\bibitem{2017NIMPA.861...28G}
A.~Galindo et~al., {\it {Calibration of a large water-Cherenkov detector at the
  Sierra Negra site of LAGO}},  {\em Nuclear Instruments and Methods in Physics
  Research A} {\bf 861} (July, 2017) 28--37.

\bibitem{ALARCON199939}
M.~Alarc\'ona et~al., {\it Calibration and monitoring of water cherenkov
  detectors with stopping and crossing muons},  {\em Nuclear Instruments and
  Methods in Physics Research Section A: Accelerators, Spectrometers, Detectors
  and Associated Equipment} {\bf 420} (1999), no.~1 39 -- 47.

\bibitem{Etchegoyen:2005xg}
{\bf Pierre Auger} Collaboration, A.~Etchegoyen, P.~Bauleo, X.~Bertou, C.~B.
  Bonifazi, A.~Filevich, M.~C. Medina, D.~G. Melo, A.~C. Rovero, A.~D.
  Supanitsky, and A.~Tamashiro, {\it {Muon-track studies in a water Cherenkov
  detector}},  {\em Nucl. Instrum. Meth.} {\bf A545} (2005) 602--612.

\end{thebibliography}\endgroup
\end{document}